\def \AAP #1 #2 {{\em Astron. Astrophys.\/} {\bf #1}, #2}
\def \AAL #1 #2 {{\em Astron. Astrophys. Lett.\/} {\bf #1}, L#2}
\def \AAR #1 #2 {{\em Astron. Astrophys. Rev.\/} {\bf #1}, #2}
\def \AAS #1 #2 {{\em Astron. Astrophys. Suppl. Ser.\/} {\bf #1}, #2}
\def \AJ #1 #2 {{\em Astron. J.\/} {\bf #1}, #2}
\def \ANNREV #1 #2 {{\em Ann. Rev. Astron. Astrophys.\/} {\bf #1}, #2}
\def \APJ #1 #2 {{\em Astrophys. J.\/} {\bf #1}, #2}
\def \APJL #1 #2 {{\em Astrophys. J. Lett.\/} {\bf #1}, L#2}
\def \APJS #1 #2 {{\em Astrophys. J. Suppl.\/} {\bf #1}, #2}
\def \APSS #1 #2 {{\em Astrophys. Space Sci.\/} {\bf #1}, #2}
\def \ASR #1 #2 {{\em Adv. Space Res.\/} {\bf #1}, #2}
\def \BAIC #1 #2 {{\em Bull. Astron. Inst. Czechosl.\/} {\bf #1}, #2}
\def \JSQRT #1 #2 {{\em J. Quant. Spectrosc. Radiat. Transfer\/} {\bf #1}, #2}
\def \MN #1 #2 {{\em Mon. Not. R. Astr. Soc.\/} {\bf #1}, #2}
\def \MEM #1 #2 {{\em Mem. R. Astr. Soc.\/} {\bf #1}, #2}
\def \PLR #1 #2 {{\em Phys. Lett. Rev.\/} {\bf #1}, #2}
\def \PASJ #1 #2 {{\em Publ. Astron. Soc. Japan\/} {\bf #1}, #2}
\def \PASP #1 #2 {{\em Publ. Astr. Soc. Pacific\/} {\bf #1}, #2}
\def \NAT #1 #2 {{\em Nature\/} {\bf #1}, #2}
\def \SAIT #1 #2 {{\em Mem.\ Soc.\ Astron.\ It.\/} {\bf #1}, #2}
\def \MESS #1 #2 {{\em The Messenger\/} {\bf #1}, #2}
\def \ASTRNACH #1 #2 {{\em Astron. Nach.\/} {\bf #1}, #2}

\def\nH{{N$_{\hbox {\footnotesize \rm H}}$}} 
\newcommand{\gtap}{\mathrel{\hbox{\rlap{\lower.55ex \hbox {$\sim$}}
                   \kern-.3em \raise.4ex \hbox{$>$}}}}
\newcommand{\ltap}{\mathrel{\hbox{\rlap{\lower.55ex \hbox {$\sim$}}
                   \kern-.3em \raise.4ex \hbox{$<$}}}}

\documentstyle{BSAXPwork}
\input epsf.sty
\begin{opening}
\title{PSR B1929$+$10 REVISITED IN X-RAYS}
\author{A. Wo\'zna$^{1,2}$, L. Kuiper$^3$, W. Hermsen$^3$}
\institute{
$^1$Max Planck Institute for Extraterrestrial Physics, Garching, Germany\\
$^2$Nicolaus Copernicus Astronomical Center, Toru$\acute{n}$,Poland\\
$^3$Space Research Organization Netherlands, Utrecht, The Netherlands}
\date{} 
\end{opening}

\begin{document}

\oddpagefooter{}{}{} 
\evenpagefooter{}{}{} 
\medskip  

\begin{abstract}
We performed  timing and spectral analysis  for PSR~B1929+10, one  of the oldest
(about $10^7$ years) of the  ordinary pulsars detected in X-rays, using archival
ROSAT, ASCA  and RXTE  data. Pulsed emission  was detected  at a more  than five
sigma level for combined ROSAT PSPC  and unpublished HRI data. Our pulse profile
is in  agreement with  that obtained by  Yancopoulos~et~al. (1994; ROSAT PSPC)
but  now with better  statistics.  An  investigation of  the behaviour of  the 
pulsed  signal as a function of energy , based on  PSPC data, provides indications 
that the pulsed fraction is changing with
energy.   The most  important new  result from  this work is derived  from the
spectral analysis. We  found that the combined ROSAT PSPC  and ASCA GIS spectrum
can satisfactorily be described by a power-law  as well as by a double black-body model
but not  by a single black-body model, as was presented  earlier by Yancopoulos~et~al.
(1994; ROSAT) and  Wang~et~al. (1997; ASCA).  Fitting  the combined  ROSAT/ASCA
0.1 - 10 keV spectrum by a power-law model we obtain a
photon index  of $2.54  \pm 0.12$  and a neutral  Hydrogen column  density (\nH)
towards  the  source of  $9.8_{-1.0}^{+1.4}  \cdot  10^{20}~\rm cm^{-2}$.   For a
double black-body fit our results are:
$T_{1} = 2.0_{-0.05}^{+0.05} \cdot 10^6~\rm K$, $T_{2} =
6.9_{-0.35}^{+0.23}  \cdot  10^6~\rm  K$,  and  \nH$=4.43_{-1.12}^{+2.08}  \cdot
10^{20}~\rm cm^{-2}$.  In  both cases the derived \nH~value  is higher than that
adopted in earlier works, but our  result is consistent with the larger distance
estimate  of  $331  \pm  10~\rm  pc$  from new  parallax  measurements  performed  by
Brisken~et~al.   (2002)  and with  the  Hydrogen  distribution  measured in  the
direction of the pulsar (Frisch \& York, 1983).  No significant pulsed signal is
found in the RXTE data.
\end{abstract}
\medskip
\section{Introduction}
PSR~B1929$+$10 is one of the closest and oldest known ordinary pulsars that
has been  detected in X-rays.  The  two latest investigations of  this object in
X-rays were  based on  ROSAT PSPC (Yancopoulos~et~al.  1994, hereafter  Y94) and
ASCA (Wang~\&~Halpern 1997, hereafter WH97) observations with exposure durations
of $45~\rm ks$ and $54~\rm ks$, respectively. Y94 obtained, after background 
subtraction, a total of $420 \pm 25$ photons in
the    0.1 - 2.0 keV    band,   corresponding    to   a    luminosity   of
$1.2\cdot10^{30}~\rm  ergs~s^{-1}$ for a  source distance  of $250~\rm  pc$, or
$\sim 3\cdot 10^{-4}$ of the  pulsar's spin-down luminosity.  After folding the
barycentered arrival  times of  the selected events  Y94 fitted the  pulse shape
using a sinusoid yielding a pulsed fraction of $0.28 \pm 0.10$. The spectrum
was fitted with  a black-body with temperature $T_\infty  \approx 3.2 \cdot 10^6~
\rm K$, which  indicated that the  implied emitting area  has a radius of  less than
$50~\rm m$. WH97 used ASCA data  from October 1994. The spectrum of PSR~B1929$+$10
over the 0.5 - 5.0~keV range was fitted with a single black-body. The neutral
Hydrogen column density was not included as  a free parameter in the fit. It was
fixed at the  level $\sim 1 \cdot 10^{20}~\rm cm^{-2}$,  and the luminosity was
calculated for a distance of $250 ~\rm pc$ used by Y94. Their results yield $T =
5.14 \pm 0.53 \cdot 10^6~\rm K$ and $L
\approx 1.28 \cdot 10^{30}~\rm ergs~s^{-1}$.  In the timing analysis
WH97 extracted events  from a circle of radius  2.5 arcmin for both GIS  2 and 3
and obtained 747 counts within this radius, 413 of which were estimated to
belong to  the background.  The lightcurve  was characterized by  a single broad
maximum, similar to the ROSAT soft X-ray profile, with a modulation significance
of  $\sim  3.1~\sigma$.   The pulse  fraction  was  estimated  to be  $0.35  \pm
0.15$. Both groups were unable  to discriminate conclusively between a thermal and
a non-thermal spectrum. While a black-body is a better overall fit, the inferred
emitting area is surprisingly small.

In this study  we revisited the ROSAT PSPC and ASCA  analyses. Moreover, we also
included ROSAT HRI  data from a combination of 3 observations  lasting 346 ks in
total,  together with  data from  a 30  ks  RXTE exposure.   The main  goal was  to
characterize the timing  and spectral properties of PSR~B1929$+$10  in more detail
using all available X-ray data.

\section{Observations}
In our study we made use of archival ROSAT PSPC/HRI, ASCA SIS/GIS and RXTE
PCA data. An overview of the X-ray observations of PSR~B1929$+$10 is presented in
Table I.

\begin{table}[!h]
\caption{X-ray observations of PSR~B1929$+$10}
\begin{center}
\begin{tabular}{lcccc}
\hline
\multicolumn{1}{l}{Instrument}  &  \multicolumn{4}{c}{ROSAT} \\  
\hline
Detector &  \multicolumn{1}{c}{PSPC}   & \multicolumn{3}{c}{HRI} \\
\hline
Obs. start & 30/03/91 & 09/10/95 & 02/10/96 & 13/04/97  \\ 
Obs. end   & 25/04/91 & 07/11/95 & 21/10/96 & 28/04/97  \\
On - Time (ks)  & 43.57 & 104.125 & 136.598 & 106.019   \\ 
 & & & & \\
\hline
Instrument  &  \multicolumn{2}{c}{ASCA} & \multicolumn{2}{c}{RXTE}  \\
\hline
Detector &  GIS  & SIS               & \multicolumn{2}{c}{PCA}   \\
\hline
Obs. start &  17/10/94 & 17/10/94 & \multicolumn{2}{c}{21/11/97} \\
Obs. end   &  19/10/94 & 19/10/94 & \multicolumn{2}{c}{21/11/97} \\
On - Time (ks) & 56.448 & 53.323 & \multicolumn{2}{c}{29.65} \\
\hline
\end{tabular}
\end{center}
\end{table}

\section{Timing analysis}
\subsection{ROSAT PSPC}
Point-like  X-ray emission was  clearly detected  in the  vicinity of  the radio
position of the  pulsar on the ROSAT PSPC Maximum Likelihood  Ratio map (MLR map,
see e.g.  Kuiper  et al.  1998). The number of source  and background counts are
derived  simultaneously by  this method.   In the  timing analysis  we extracted
events within an optimum radius (the radius where the signal-to-noise ratio S/N 
maximizes) of $50^{\prime \prime}$ from the X-ray  centroid. The 604 events 
(0.1  - 2.5 keV) falling  within this radius
have  subsequently  been  barycentered  and folded  through  appropriate  pulsar
parameters (see e.g. Table  II for general characteristics of  PSR~B1929$+$10) to obtain
the pulse-phase distribution shown in left panel of Fig. 1.  The profile is
characterized  by one  broad  maximum  and deviates  from  a statistically  flat
distribution at a $4.2~\sigma$ level when applying a $Z_{1}^{1}$ method (Buccheri et
al. 1983). Due to the limited accuracy of the ROSAT absolute timing we could not
relate the  X-ray phase  to the radio  phase.  In  order to quantify  the pulsed
fraction we have  applied two methods: one based  on bootstrapping (Swanepoel et
al. 1996) and another based  on sinusoid fitting.  For broad sinusoidal profiles
it is expected  that the bootstrap method will  systematically underestimate the
genuine pulsed fraction  because parts of the leading and  trailing wings of the
broad profile are  considered to be part of the  interval used to
construct the unpulsed level.  Applying  the bootstrap method we obtain a pulsed
fraction, defined  as $f_{pulsed} =  \frac{N_{P}}{N_{P} + N_{DC}}$, of  $0.24 \pm
0.07$.  The number of pulsed  counts $N_{P}$ follows directly from the bootstrap
method, while  the number of  DC counts $N_{DC}$  requires an estimate  of the
number of background  counts within the extraction radius of  $50^{\prime \prime}$.  
From the MLR map we could estimate that $140.5 \pm 2.2$ background counts are 
expected within the extraction radius.
\begin{table}[h]
\caption{Pulsar characteristics$^a$}
\begin{center}
\begin{tabular}{ll}
\hline
\multicolumn{1}{c}{Parameter} & \multicolumn{1}{c}{Value} \\
\hline
PSR                       & B1929$+$10 $\equiv$ J1932+1059 \\
Right Ascension (J2000)   & $\rm 19^{h}~32^{m}~13^{s}.899$ \\
Declination (J2000)       & $\rm 10^{\circ}~59'~31''.99$ \\
Period                    & 0.226518$\rm~s$ \\
Period derivative         & 1.15661~$\cdot~10^{-15}$ \\
Age                       & 3.1 $\rm~Myr$ \\
Magnetic field$^b$        & 0.51~$\cdot~10^{12} \rm~G$ \\
Distance$^c$              & 331$\rm~pc$ \\
Spin--down luminosity$^d$ & 3.89~$\cdot~10^{33} \rm~ergs~s^{-1}$ \\ 
\hline
\multicolumn{2}{l}{\footnotesize{$^a$ Taylor et al. 1993.}} \\
\multicolumn{2}{l}{\footnotesize{$^b~B = (3Ic^3P\dot{P}/8\pi^2R^6)^{1/2}$,
for $R = 10^6\rm~cm$, $I = 10^{45}\rm~g~cm^2$.}} \\
\multicolumn{2}{l}{\footnotesize{$^c$ From parallax measurements by
Brisken et al. 2002.}} \\
\multicolumn{2}{l}{\footnotesize{$^d~\dot{E} = 4\pi^2I\dot{P}P^{-3}$,
for $I = 10^{45}\rm~g~cm^2$.}} \\
\end{tabular}
\end{center}
\end{table}

Because the pulse profile shows one broad maximum sinusoid fitting using one
harmonic  provides  an  adequate   description  of  the  measured  distribution.
Applying sinusoid fitting to the 0.1 - 2.5~keV PSPC pulse profile (see Fig. 2
\emph{left panel}) we find a pulse fraction of $0.36 \pm 0.08$ slightly higher than the 
bootstrap value, as expected, and consistent with the value found earlier by Y94.

By selecting the  events further on energy we applied  the same sinusoid fitting
method to  the 0.1  - 1.28~keV  and 1.28  - 2.5~keV  pulse profiles.   These two
profiles  are also  shown  in Fig.   1:  0.1 -  1.28~keV, \emph{middle  panel};
modulation significance  $3.0~\sigma$, and 1.28  - 2.5~keV,  \emph{right panel},
modulation  significance  also $3.0~\sigma$.   The  pulsed  fractions applying 
sinusoid fitting for these energy ranges are $0.33 \pm 0.09$ and $0.42 \pm 0.13$, 
respectively.

\begin{figure}[h]
\begin{minipage}[t]{4.35cm}
\epsfysize=4.35cm \epsfxsize=4.35cm \hspace{0.0cm} \vspace{0.0cm}
\epsfbox{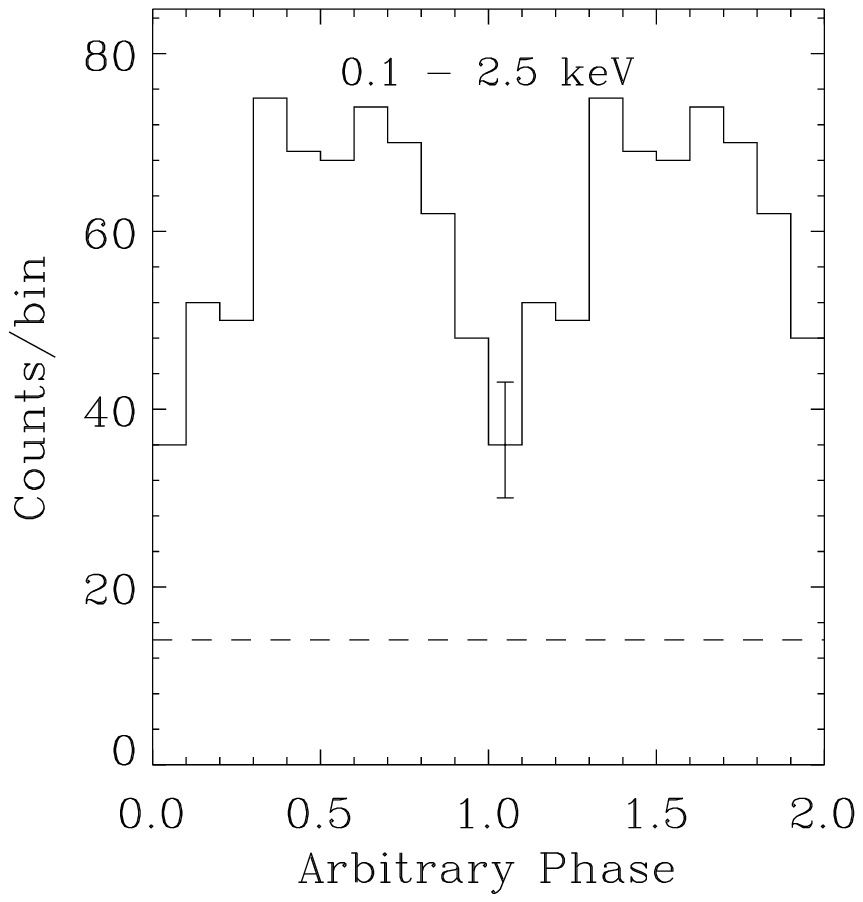}
\end{minipage} \hskip0cm \begin{minipage}[t]{4.35cm}
\epsfysize=4.35cm \epsfxsize=4.35cm \hspace{0.0cm} \vspace{0.0cm}
\epsfbox{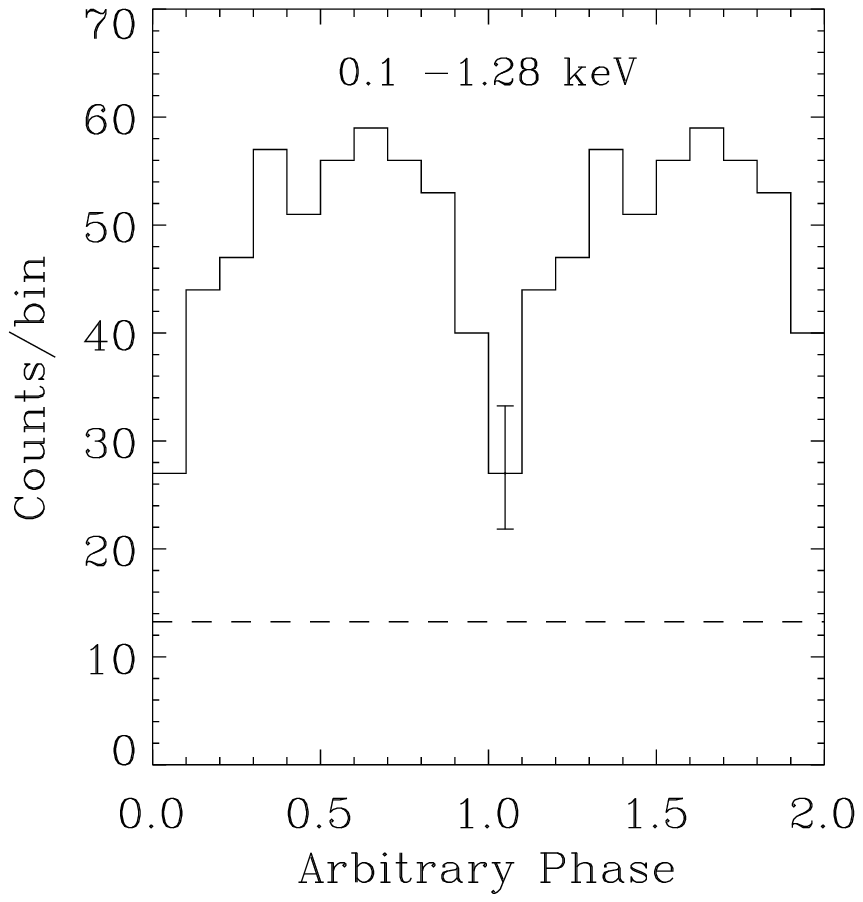}
\end{minipage} \hskip0cm \begin{minipage}[t]{4.35cm}
\epsfysize=4.35cm \epsfxsize=4.35cm \hspace{0.0cm} \vspace{0.0cm}
\epsfbox{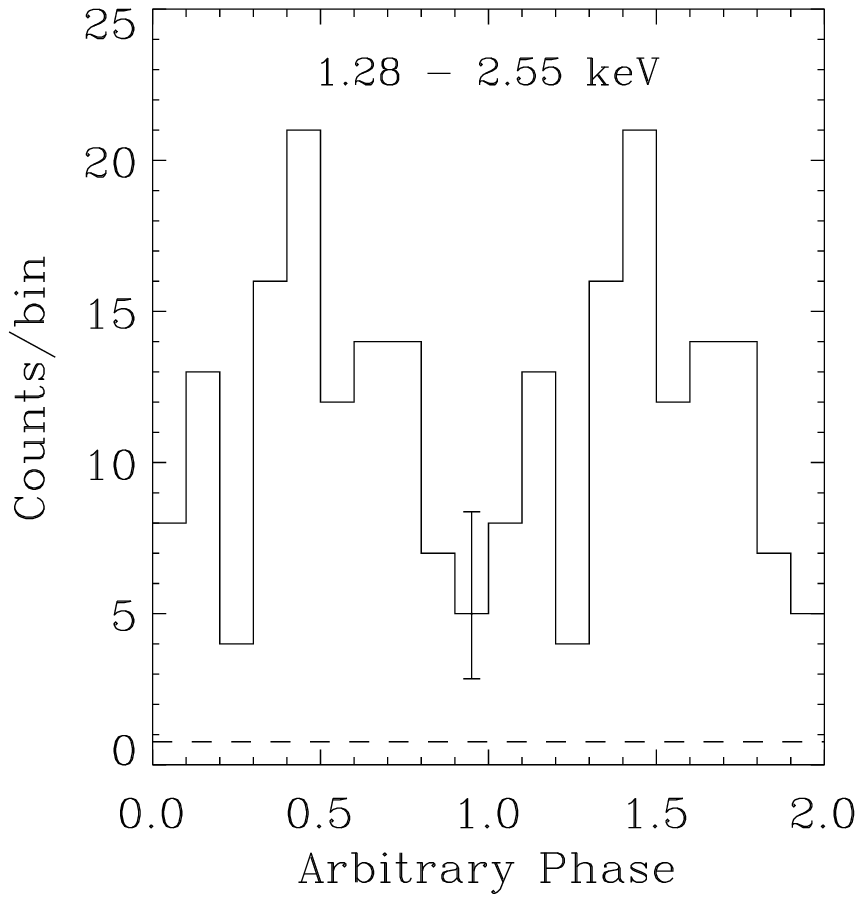}
\end{minipage}
\caption{PSR~B1929$+$10 phase histograms from ROSAT PSPC data in three energy ranges,
\emph{left panel}: 0.1 - 2.5~keV, \emph{center panel}: 0.1 - 1.28~keV,
\emph{right panel}:  1.28 - 2.5~keV. Two cycles  are shown for  clarity. A
typical error bar is shown on each panel. The significance for a deviation 
from a statistically flat distribution is $4.15~\sigma$ (604 events), 
$3.0~\sigma$ (490 events), $3.0~\sigma$ (114 events), respectively.}
\end{figure}

\subsection{ROSAT HRI}
For  the  3 long  ROSAT  HRI  observations (see  Table  I)  the timing  analysis
consisted in  extracting the events from  a $11^{\prime \prime}$ aperture around
the centroid of
the  PSR~B1929$+$10 counterpart  and subsequently  folding the  barycentered event
times using  an appropriate  pulsar ephemeris. The  obtained pulse  profiles for
each observation are all of low significance (between 2 - 3 sigma), although the
broad  bump can be  discerned easily.  Cross-correlating each  of the  three HRI
profiles with  the 0.1 -  2.5~keV  PSPC profile and  applying the shifts  in the
combination of the three separate  aligned profiles, yielded a phase distribution
which deviates from a flat distribution at the $3.7~\sigma$ level (see Fig.~2
\emph{middle panel}). In this case the sinusoid fitting method yielded a pulsed 
fraction  of $0.21  \pm 0.05$,  lower but  consistent with  the PSPC  value. The
combined ROSAT PSPC  and HRI profile is shown in the  \emph{right panel} of Fig.~2
and it deviates from being  flat at a $5.6~\sigma$ level.  The pulsed fraction
of the combined profile is $0.25 \pm 0.04$.
\begin{figure}
\begin{minipage}[t]{4.35cm}
\epsfysize=4.35cm \epsfxsize=4.35cm \hspace{0.0cm} \vspace{0.0cm}
\epsfbox{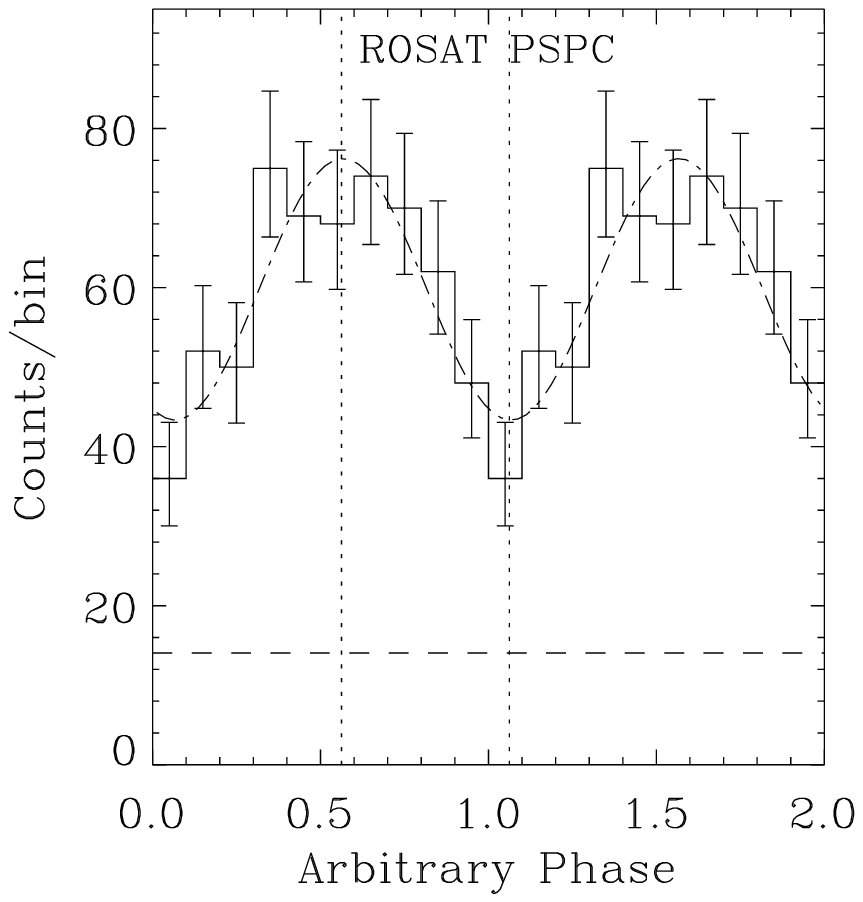}
\end{minipage} \hskip0cm \begin{minipage}[t]{4.35cm}
\epsfysize=4.35cm \epsfxsize=4.35cm \hspace{0.0cm} \vspace{0.0cm}
\epsfbox{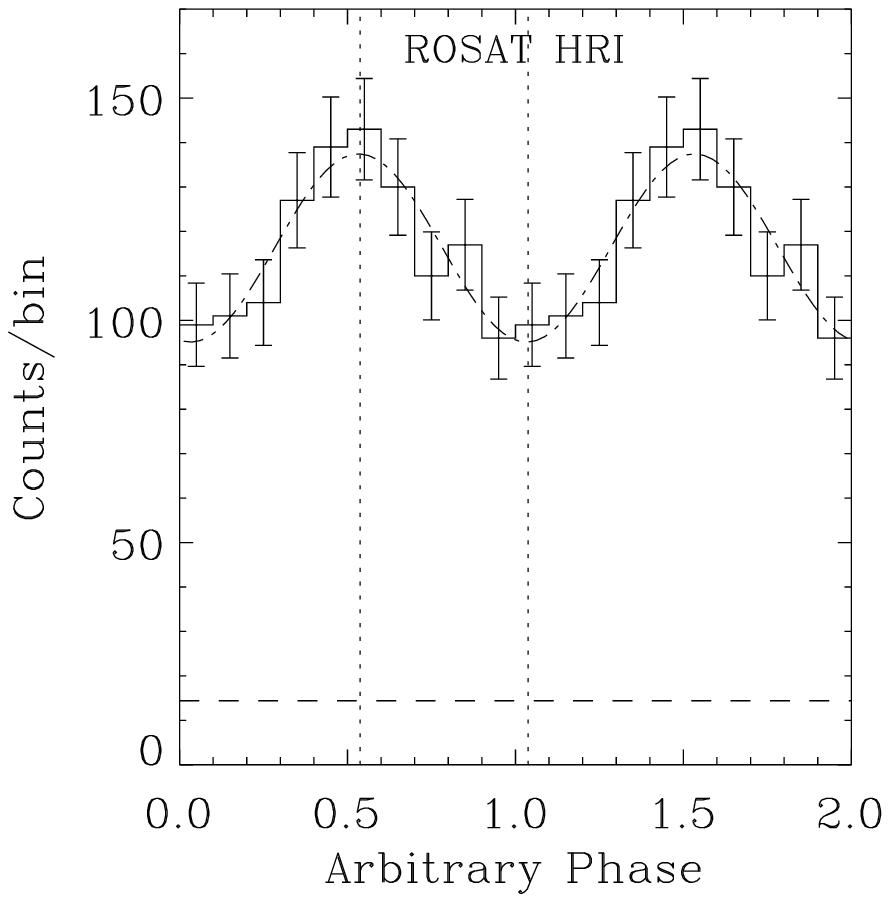}
\end{minipage} \hskip0cm \begin{minipage}[t]{4.35cm}
\epsfysize=4.35cm \epsfxsize=4.35cm \hspace{0.0cm} \vspace{0.0cm}
\epsfbox{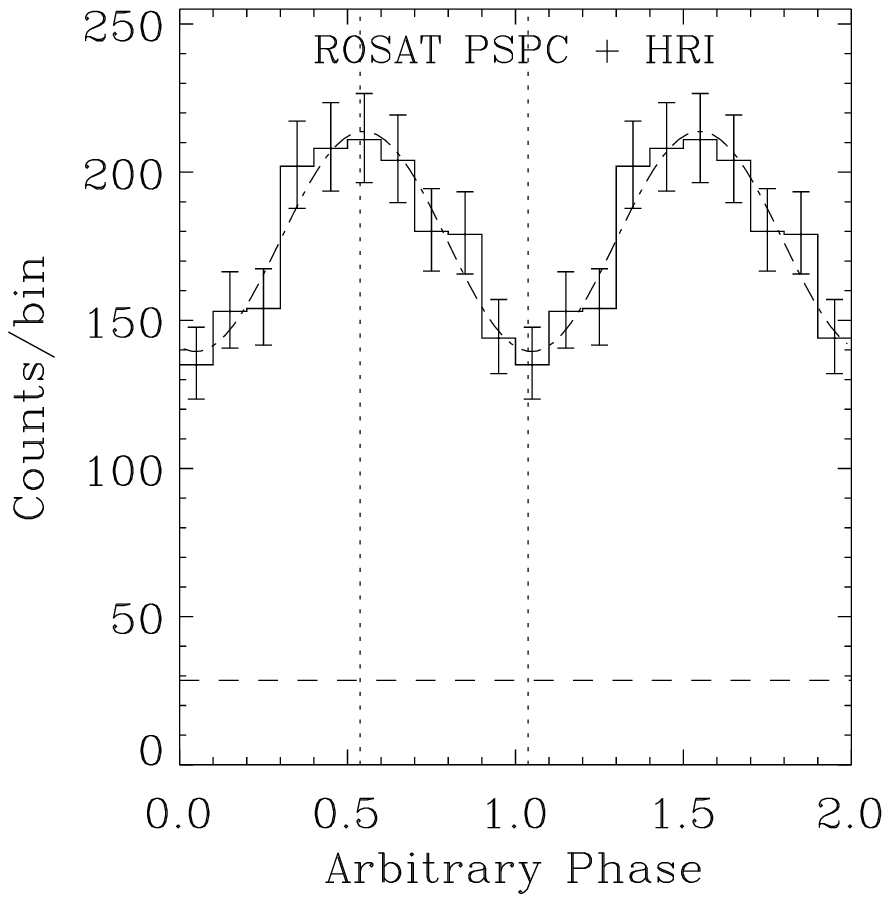}
\end{minipage}
\caption{Light curves for the ROSAT PSPC and HRI observations of PSR~B1929$+$10. 
\emph{Left panel}: only PSPC (0.1 - 2.5~keV), \emph{center panel}: only 
HRI (0.1  - 2.4~keV),  \emph{right panel}: combined  PSPC and HRI. Two 
cycles are shown  for clarity. Error bars are shown in  each panel.  The 
significance  for a  deviation from  a  statistically flat distribution     
is    $4.15~\sigma$,    $3.7~\sigma$     and    $5.6~\sigma$,
respectively.  The  dashed  horizontal   lines  indicate  the  background  level
determined from  a spatial analysis.  The vertical dotted lines  correspond to
minimum and maximum phase of the sinusoid function
fit (dot - dashed curve).}
\end{figure}

\subsection{ASCA GIS}
Of the two  detector systems aboard ASCA only the  GIS detectors have sufficient
timing resolution  to support timing  analyses at millisecond scales.  Data from
the high and medium rate telemetry  mode were used. The 3.9 ms timing resolution
valid for these rates was sufficiently high to sample the pulse period of 227 ms
(see Table II).  In the spatial  analysis a Maximum Likelihood method similar to
the one used for the ROSAT  PSPC/HRI data, provided the optimum position for the
X-ray  counterpart  of  PSR~B1929$+$10.  An extraction  radius  of  $130^{\prime
\prime}$ from the  optimum position was found to optimize the  S/N ratio. In the
0.5  -  5~keV   energy  interval  493  event  timetags   were  barycentered  and
subsequently  folded   through  appropriate  timing  parameters   to  yield  the
corresponding  pulse profile.  This 0.5  -  5~keV profile  is shown  in Fig.   3
\emph{left panel} and deviates from uniformity at a $2.7~\sigma$ level. The same
broad enhancement is visible as in  the ROSAT profiles at softer X-ray energies.
Constraining the  energy to  the 0.5 -  2.0~keV interval, fully  overlapping the
ROSAT  energy  window,  yielded  a  significance of  $3.1~\sigma$  (see  Fig.  3
\emph{right panel}).  The pulsed fractions derived from sinusoid fitting for the
(integral) 0.5 -  5~keV and (ROSAT overlapping) 0.5 -  2.0~keV energy ranges are
$0.36  \pm  0.11$  and  $0.69  \pm  0.18$, respectively.  The  former  value  is
consistent with  the one found  by W97, while  the latter value derived  for the
ROSAT overlapping  interval seems to confirm  the increasing trend  seen at soft
X-ray energies for the pulsed fraction as a function of energy.
\begin{figure}
\begin{center}
\begin{minipage}[t]{4.35cm}
\epsfysize=4.35cm \epsfxsize=4.35cm \hspace{0.0cm} \vspace{0.0cm}
\epsfbox{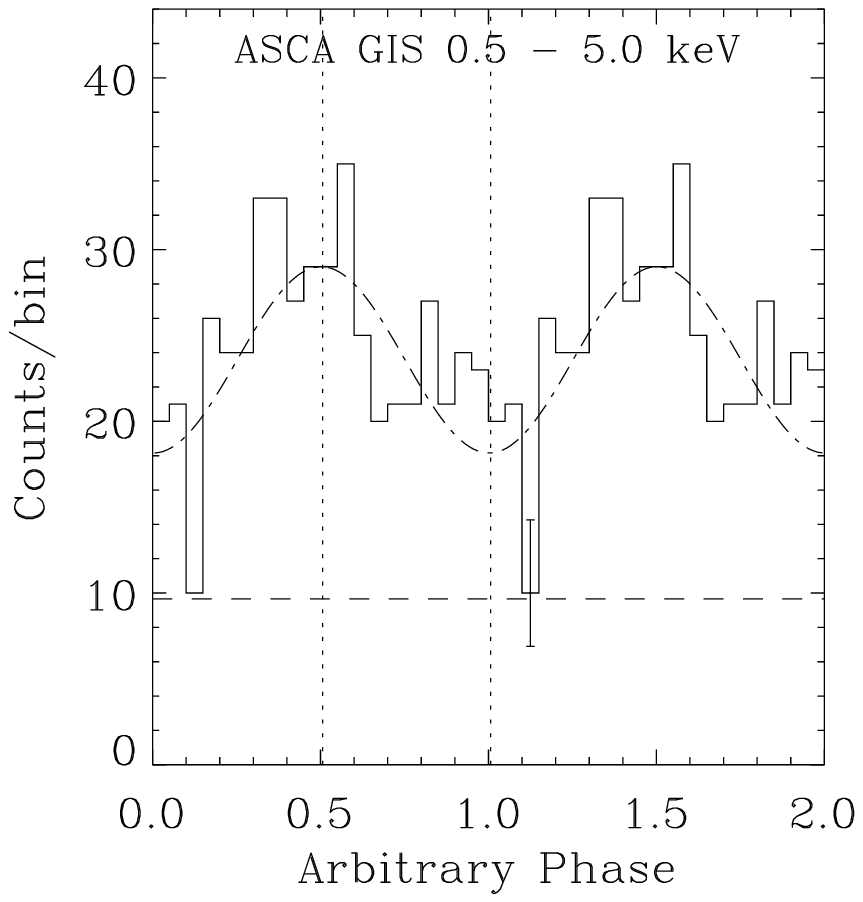}
\end{minipage} \hskip0cm \begin{minipage}[t]{4.35cm}
\epsfysize=4.35cm \epsfxsize=4.35cm \hspace{0.0cm} \vspace{0.0cm}
\epsfbox{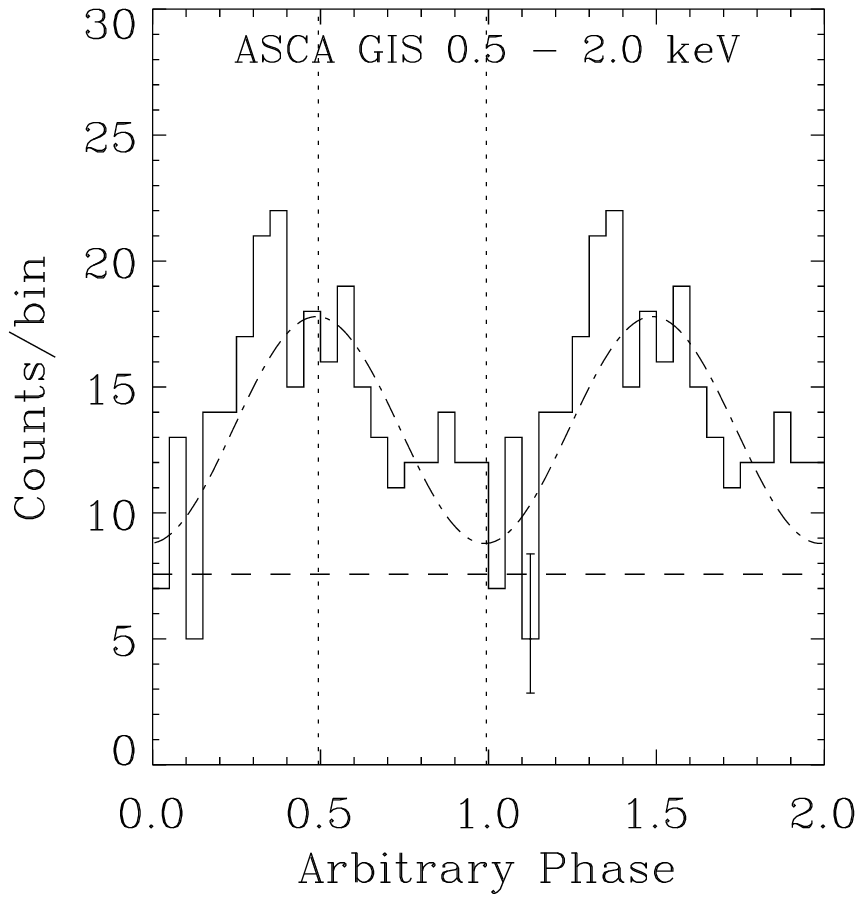}
\end{minipage}
\end{center}
\caption{The ASCA profiles of PSR~B1929$+$10 in two energy intervals:
\emph{left panel} 0.5 - 5.0~keV (significance $2.7~\sigma$) and \emph{right panel}
0.5 - 2.0~keV (significance $3.1~\sigma$).  The dot~-~dashed curves indicate the
sinusoid function fitted to the data, while the vertical dotted lines correspond
to the minimum  and maximum phase of these fits.  In  both panels the horizontal
dashed lines indicate the background  levels determined from a spatial analysis.
A typical error bar is shown in each panel. Two cycles are shown for clarity.}
\end{figure}

\subsection{RXTE PCA}
Finally, we analysed RXTE PCA (2 - 60~keV) data from a 30 ks observation of PSR
B1929$+$10  (Obs.  id.   20156; 21-Nov-1997)  obtained  in Good  Xenon mode,  time
tagging each event with a $0.9~\rm \mu s$ time resolution. For the timing analysis we
used {\sf ftools}  program {\sf fasebin} to select events from  the top layer of
each involved PCU, to barycenter the times of the selected events and finally to
fold the barycentered  event times with an appropriate  ephemeris. The ephemeris
we used in  the folding procedure within {\sf fasebin} is  based on radio timing
observations  of PSR~B1929$+$10  with  the 32m  TCfA  radio-telescope at  Toru\'n,
Poland since the  beginning of June 1997. A  dual-channel, circular polarization
L-band  receiving system  at  frequencies around  1.7~GHz  and a  2~x~64~x~3~MHz
pulsar back-end,  the Penn Stata Pulsar  Machine~-~2 (Konacki et  al.  1999) are
used.   For analyzing  the radio  pulsar timing  data the  {\sf  TEMPO} software
package ({\sl http://pulsar.princeton.edu/tempo}) has been used.

The  folding  procedure  embedded  in  {\sf fasebin}  resulted  in  pulse  phase
histograms for each  of the 256 PHA channels. Unfortunately, in  none of the PHA
channels or channel intervals significant pulsed emission has been detected.

\section{Spectral analysis}
In the spectral  analysis we first derived the number of  counts assigned to the
X-ray  counterpart of  PSR~B1929$+$10  in  several narrow  energy  slices using  a
spatial (MLR) analysis  which took into account the  presence of nearby sources.
We applied this approach to the ROSAT PSPC-B (20 energy bins in the interval 0.1
- 2.5~keV  ) and  ASCA GIS 2+3  (10 energy bins  in the  interval 0.5 -  10~keV)
data. We  chose for ASCA GIS data  (contrary to W97) because the GIS sensitivity is
higher than the SIS sensitivity for energies  beyond $\sim 5~\rm keV$ and because of
uncertainties in the SIS efficiency for energies below 1~keV.  In the combined 
ROSAT PSPC-B and ASCA GIS 2+3 spectral fits we used  the latest upgrades for the 
response matrices and we took  into account the vignetting  corrections for the  
off-axis (3.7 arcmin)
ASCA observation. 
\begin{figure}[h]
\begin{center}
\begin{minipage}[t]{6.5cm}
\epsfysize=6cm \epsfxsize=6.5cm \hspace{0.0cm} \vspace{0.0cm}
\epsfbox{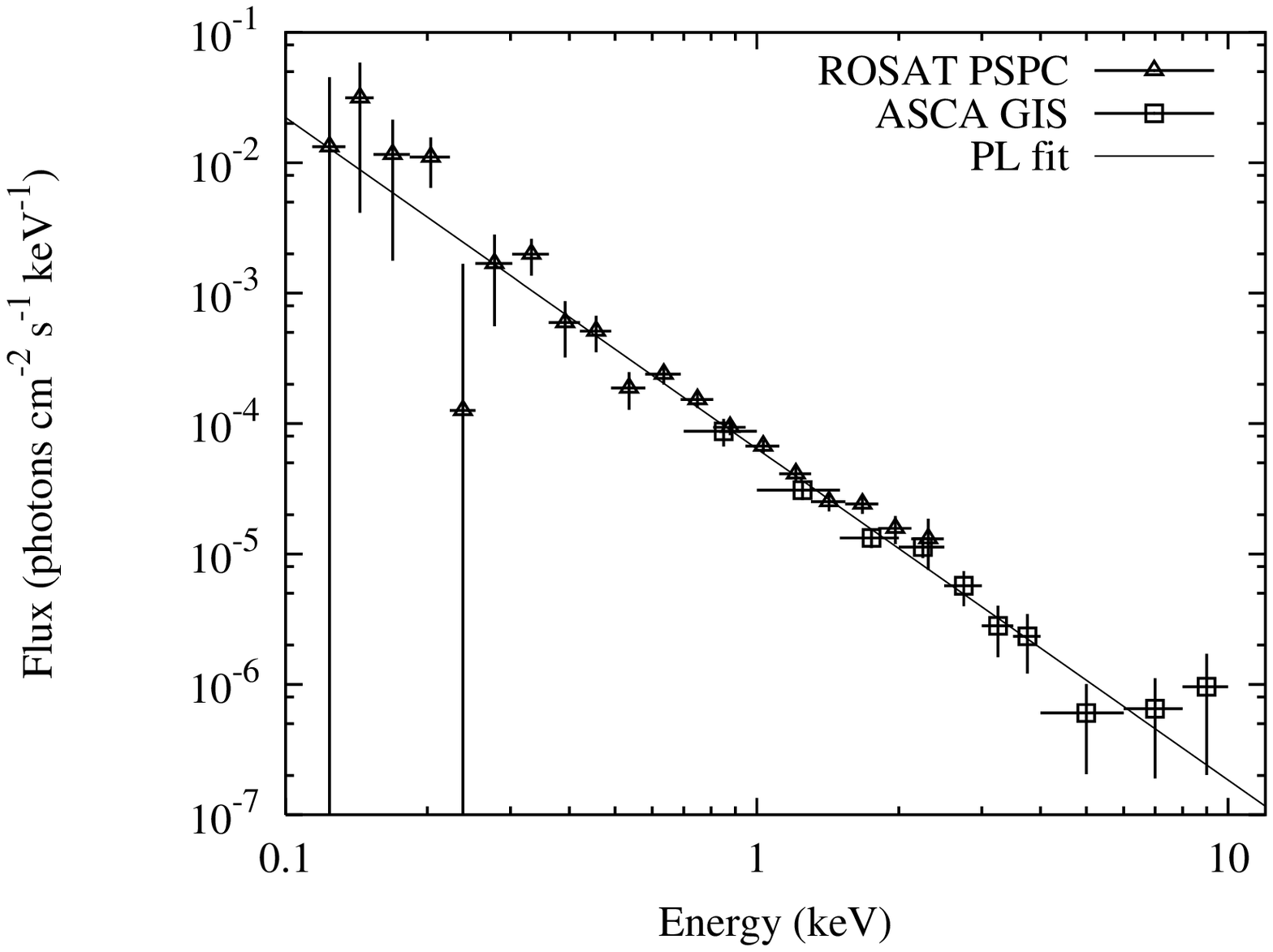}
\end{minipage} \hskip0cm \begin{minipage}[t]{6.5cm}
\epsfysize=6cm \epsfxsize=6.5cm \hspace{0.0cm} \vspace{0.0cm}
\epsfbox{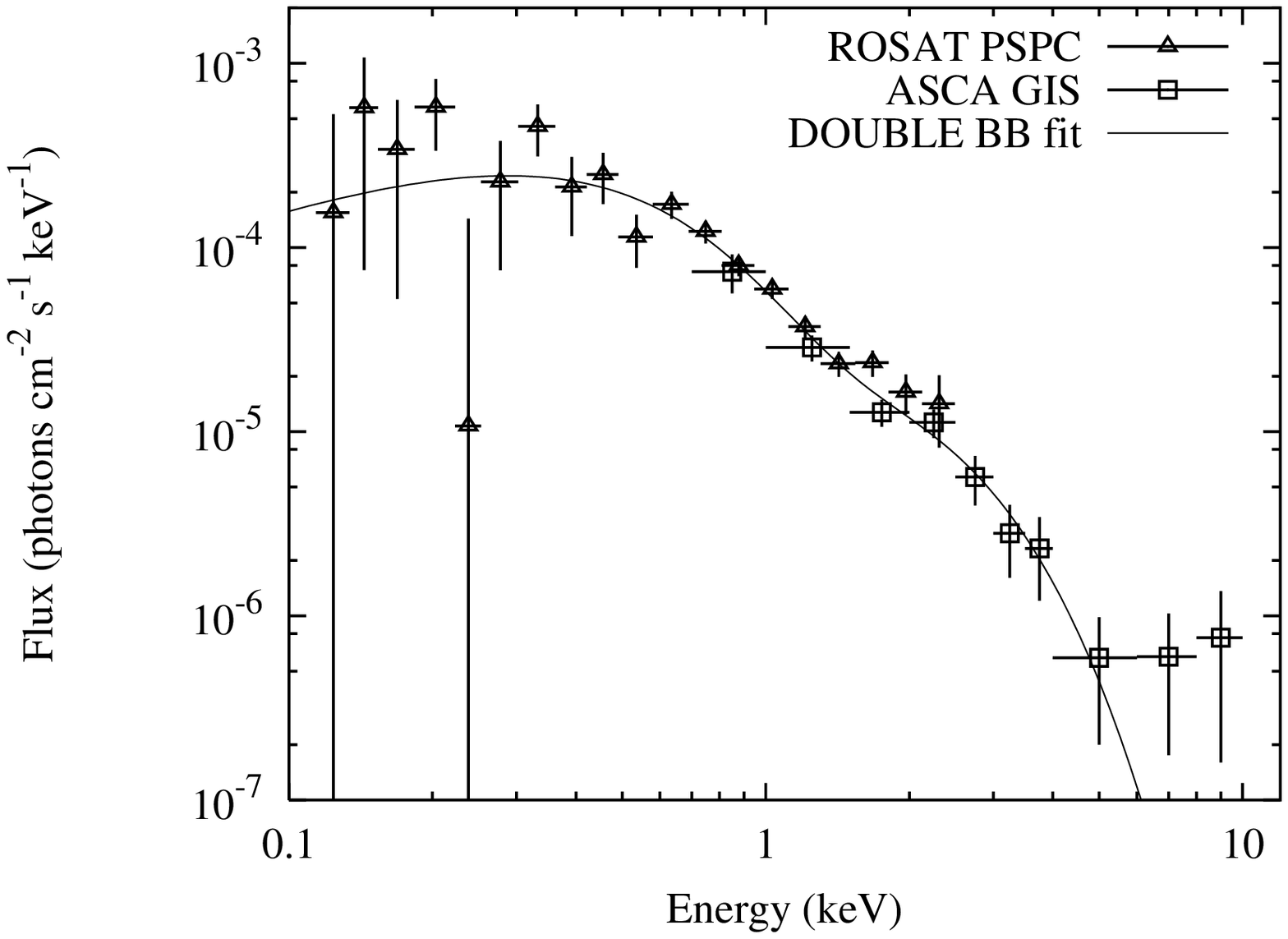}
\end{minipage}
\end{center}
\caption{0.1 - 10~keV spectrum of PSR~B1929$+$10 from combined ROSAT PSPC-B and 
ASCA GIS 2+3 data: \emph{left panel} power-law model, \emph{right panel} double 
black-body model.}
\end{figure}
The spectral models we fitted to the combined data set consist
of  a single  absorbed  power-law,  a single  absorbed  black-body, an  absorbed
black-body plus power-law and an  absorbed double black-body. The best fits were
obtained  for a  single absorbed  power-law (Fig.~4 \emph{left panel})
and an  absorbed double black-body model (Fig.~4 \emph{right panel}). The fit 
results for the  latter two cases, each with a free Hydrogen column density, 
are shown in Table III.
\begin{table}[h]
\caption{Fits to combined ROSAT and ASCA spectrum in the energy range 0.1 - 10~keV}
\begin{center}
\begin{tabular}{l l |l l}
\hline
Parameter & Power Law  Fit & Parameter & Double Black-Body Fit \\
\hline
$C^{a}$ & $6.41^{+0.4}_{-0.4} \cdot 10^{-5}$  &
$C_{1}$ & $0.0117^{+0.0017}_{-0.0018}$ \\
$\alpha$ & $2.54^{+0.12}_{-0.13}$ &
$kT_{1}$ & $0.176^{+0.004}_{-0.004}~\rm keV$ \\
$N_H$ & $9.8^{+1.4}_{-1.0}\cdot 10^{20}~\rm cm^{-2}$ &
$C_{2}$ & $7.879^{+1.32}_{-1.26} \cdot 10^{-5}$ \\
$\chi^{2}_{\nu}/\nu$ & 1.146/27 &
$kT_{2}$ & $0.595^{+0.02}_{-0.03}~\rm keV$ \\
 & & $N_H$ & $4.43^{+2.08}_{-1.12} \cdot 10^{20}~\rm cm^{-2}$ \\
 & & $\chi^{2}_{\nu}/\nu$ & 1.132/25 \\
\hline
\multicolumn{4}{l}{\footnotesize{$^a$ Normalization at 1 keV 
$\rm (ph~cm^{-2}~s^{-1}~keV^{-1})$.}}
\end{tabular}
\end{center}
\end{table}

In case  of the double  black-body model we are dealing with two different
thermal X-ray  components with temperatures $(2.0_{-0.05}^{+0.05}) \cdot  10^{6} ~\rm K$
and  $(6.9_{-0.35}^{+0.23}) \cdot  10^{6}~\rm   K$  for  $T_{1}$   and  $T_{2}$,
respectively.  For the characteristic age of PSR~B1929$+$10 ($\tau \simeq 3.1~\rm Myr$)
cooling models  predict a  surface temperature  of the neutron  star to  be $\ltap
10^{5}~\rm K$, much too low to be responsible for the observed X-ray spectrum of
PSR~B1929$+$10. However, a plausible explanation could be that the observed X-ray 
radiation originates from a heated polar cap characterized by two different thermal 
components (see e.g. Cheng \& Zhang 1999). 
The  \nH~value obtained from the fit is about two times higher than that used by 
Y94 and WH97.  Moreover, the (non-thermal) power-law fit resulting in a photon index 
of $2.54 \pm 0.12$  yielded an estimate for the neutral Hydrogen density  towards the source of 
$9.8^{+1.4}_{-1.0} \cdot 10^{20} \rm  cm^{-2}$. This value is six  times larger 
than the  adopted one in earlier works (Y94, WH97). In both  cases the derived column 
density is  consistent with the updated (radio)  distance determined  by Brisken  et al.   
(2002), and  the Hydrogen distribution  measured  in  the  direction   of  the  
pulsar  (Frisch  \&  York, 1983). Additionally, as Pavlov et  al.  (1996) mentioned, 
values of \nH~obtained from direct measurements of stars in  the neighbourhood of 
the pulsar are rather controversial.  They differ  from $< 10^{19}$ up to  
$10^{21}~\rm cm^{-2}$. This suggests that interstellar medium is very patchy in that 
direction.

\section{Conclusion}

Our revisited analysis of X-ray data of PSR 1929$+$10 from ROSAT, ASCA and RXTE 
observations yielded an improved timing signal at soft X-rays, $5.6~\sigma$ for 
the combined ROSAT PSPC/HRI pulse profile, and a non-detection for energies
beyond $\sim 2$~keV (RXTE PCA). Analyzing the pulse profile at soft X-rays (ROSAT PSPC)
in two different energy bands suggested that the pulsed fraction increases from 
$0.33 \pm 0.09$ in the 0.1~-~1.28~keV band to $0.42 \pm 0.13$ in the 1.28~-~2.5~keV
band (even $0.69 \pm 0.18$ for the 0.5~-~2~keV ASCA GIS band). 
This tendency must stop abruptly near $\sim 2$ keV to be
consistent with the non-detection of pulsed emission above $\sim 2$~keV.
A spectral analysis of the combined ROSAT/ASCA data in the 0.1 - 10~keV energy 
range indicated that both an absorbed power-law (non-thermal) model and an absorbed double 
black-body model provide an adequate description of the observed X-ray spectrum.
The derived Hydrogen column density \nH~ is in both cases much larger than the value
used in previous X-ray studies. However, the newly determined \nH~ values are more consistent
with the updated (increased) distance to PSR~B1929$+$10 in combination with the
measured Hydrogen distribution in the direction of the pulsar. 

A scheduled observation with the XMM-Newton  satellite could confirm our results 
and the much better statistics would help  to fix the column density \nH~ and
to discriminate conclusively between thermal and non-thermal models for the X-ray 
emission from this pulsar.

\acknowledgements
This research was supported by KBN grant 2P03D02117 and Polish Foundation
of Astronomy. Aga Wo\'zna acknowledges people from SRON for their
hospitality during her stay there, when most of this work was done.

\end{document}